# Artificial Synaptic Arrays Intercoupled by Nanogranular Proton Conductors for Building Neuromorphic Systems

Changjin Wan, Guodong Wu, Liqiang Guo, Liqiang Zhu, and Qing Wan

Ningbo Institute of Materials Technology and Engineering, Chinese Academy of Sciences, Ningbo 315201, People's Republic of China

**The highly parallel process in the neuron networks is mediated through a mass of synaptic interconnections. Mimicking single synapse behaviors and highly paralleled neural networks has become more and more fascinating and important. Here, oxide-based artificial synaptic arrays are fabricated on P-doped nanogranular $SiO_2$-based proton conducting films at room temperature. Synaptic plasticity is demonstrated on individual artificial synapse. Most importantly, without any intentional hard-wired connection, such synaptic arrays are intercoupled due to the electric-field induced lateral proton modulation. The natural interconnection is weakly correlative with distance, and is important for neural networks. At last, paralleled summation is also mimicked, which provides a novel approach for building future brain-like computational systems.**

**Corresponding author: Qing Wan**

**Email: wanqing@nimte.ac.cn**

The human brain, with ~$10^{11}$ neurons and ~$10^{15}$ synapses [1], can act more robust, plastic, fault-tolerant and much lower energy consumption than any current electric computer. For example, it can simultaneously gather thousands of sensory inputs and interpret them in real time as a whole and react appropriately, abstracting, learning, planning and inventing.[2] Its high efficiency is due to the parallel process mode, and the highly parallel process is through numerous connectivity between neurons (~$10^4$ in a mammalian cortex) and the plastic synapses.[3] Synapses are also crucial to biological computations that underlie perception and learning.[4] Therefore, mimicking plastic synapse or even the highly paralleled neural networks has attracted considerable attention.[5-14]

Synapse is essential to brain function, and it allows a neuron to pass an electric or chemical signal to another neuron. In the beginning, CMOS based architectures have been designed to emulate behaviors of synapse, but this approach requires many devices and high power consumption.[5-7] Thus the development of new hardware devices having a brain-inspired massively parallel, dynamical architecture and radically different from contemporary IT technology is today a top objective for scientists. Recently, several novel devices, such as $Ag_2S$ atomic switch, nanoscale memristor and phase change memory cell were reported to mimic biological synapse.[8-14] However, as Christian K. Machens mentioned, simply "building" a brain from the bottom up has a shortcoming that connected by hard-wire with low flexible would fail to capture its essential function–complex behavior.[15] Motivated by this challenge, we report the fabrication of oxide-based artificial synaptic arrays on phosphorus-doped nanogranular $SiO_2$-based proton conductor films. It is interesting to find that without any intentional hard-wire connection, such synaptic arrays are intercoupled due to the lateral electric-field induced proton migration. Such natural interconnection is weakly correlative with distance, and paralleled summation function is also experimental demonstrated.

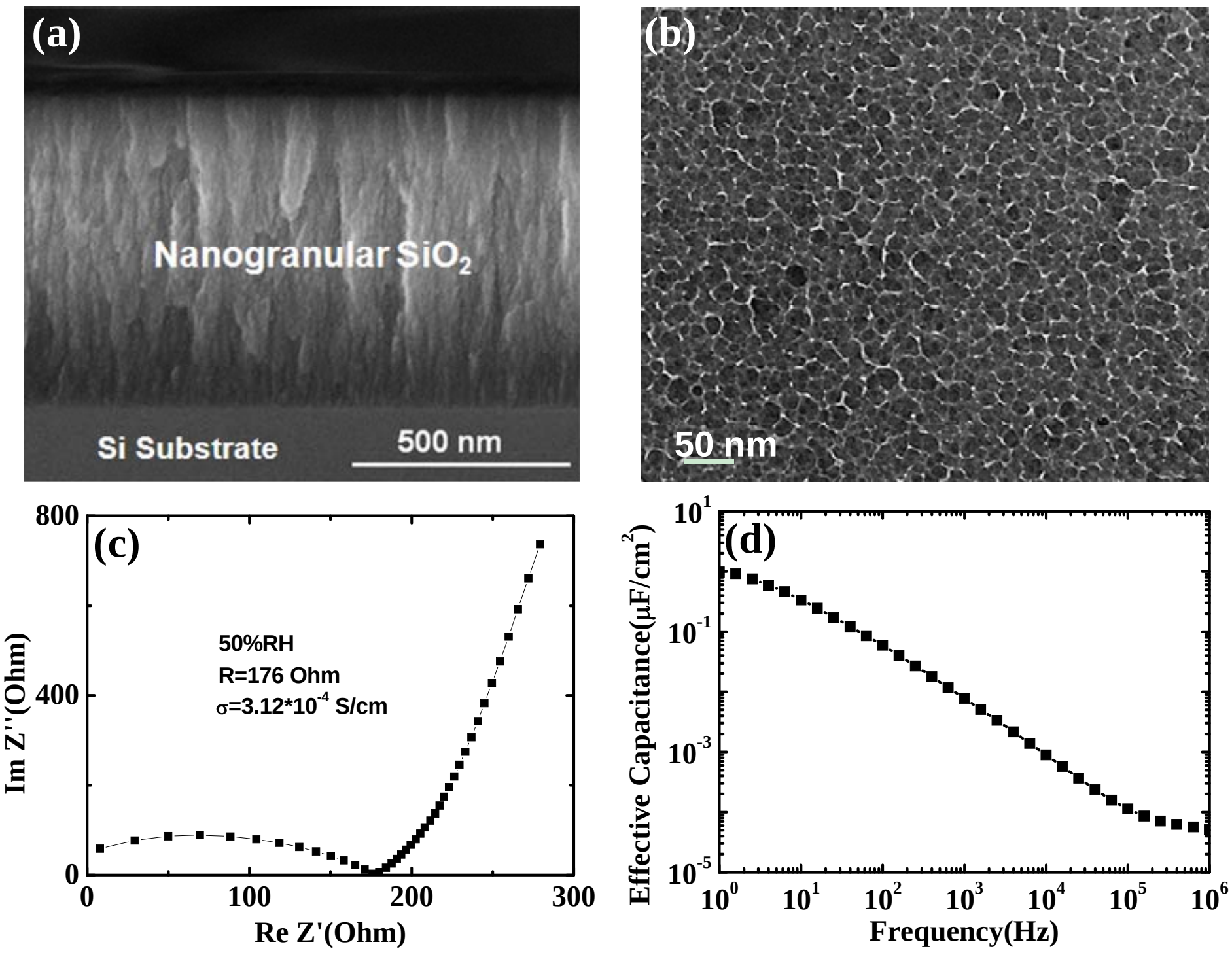

Figure 1. (a) Cross-section SEM image of the P-doped $SiO_2$ nanogranular film on silicon (100) substrate. (b) TEM images of P-doped $SiO_2$ nanogranular film deposited on TEM Cu grid. (c) Cole-Cole plots of proton conductive P-doped $SiO_2$ nanogranular film. (d) Frequency dependent lateral capacitance of the nanogranular $SiO_2$ film measured from two electrodes of $S_1$ and $S_2$.

---

As shown in Fig. 1 (a), the thickness of the P-doped $SiO_2$ nanogranular film is estimated to be ～680 nm by scanning electron microscopy (SEM) image. P-doped $SiO_2$ film deposited by PECVD method at room temperature has a nanogranular structure with the mean particle size of 8 nm, as shown by the transmission electron microscopy (TEM) image in Fig. 1 (b). As shown in Fig. 1 (c), such nanogranular $SiO_2$ film shows a high proton conductivity of $3.1\times10^{-4}$ S/cm at room temperature with the relative humidity of 60%. Figure 1 (d) shows the frequency dependent lateral capacitance of the nanogranular P-doped $SiO_2$ film measured from two electrodes of $S_1$ and $S_2$. The high specific capacitance (~1.0 μF/cm$^2$) at low frequency is due to the electric-double-layer (EDL) effect. For neural network emulate, the proton in the $SiO_2$ nanogranular film can analogous to the neurotransmitter and electron concentration in the patterned IZO film can stand for the weight (or retention) of the 'synapse'.

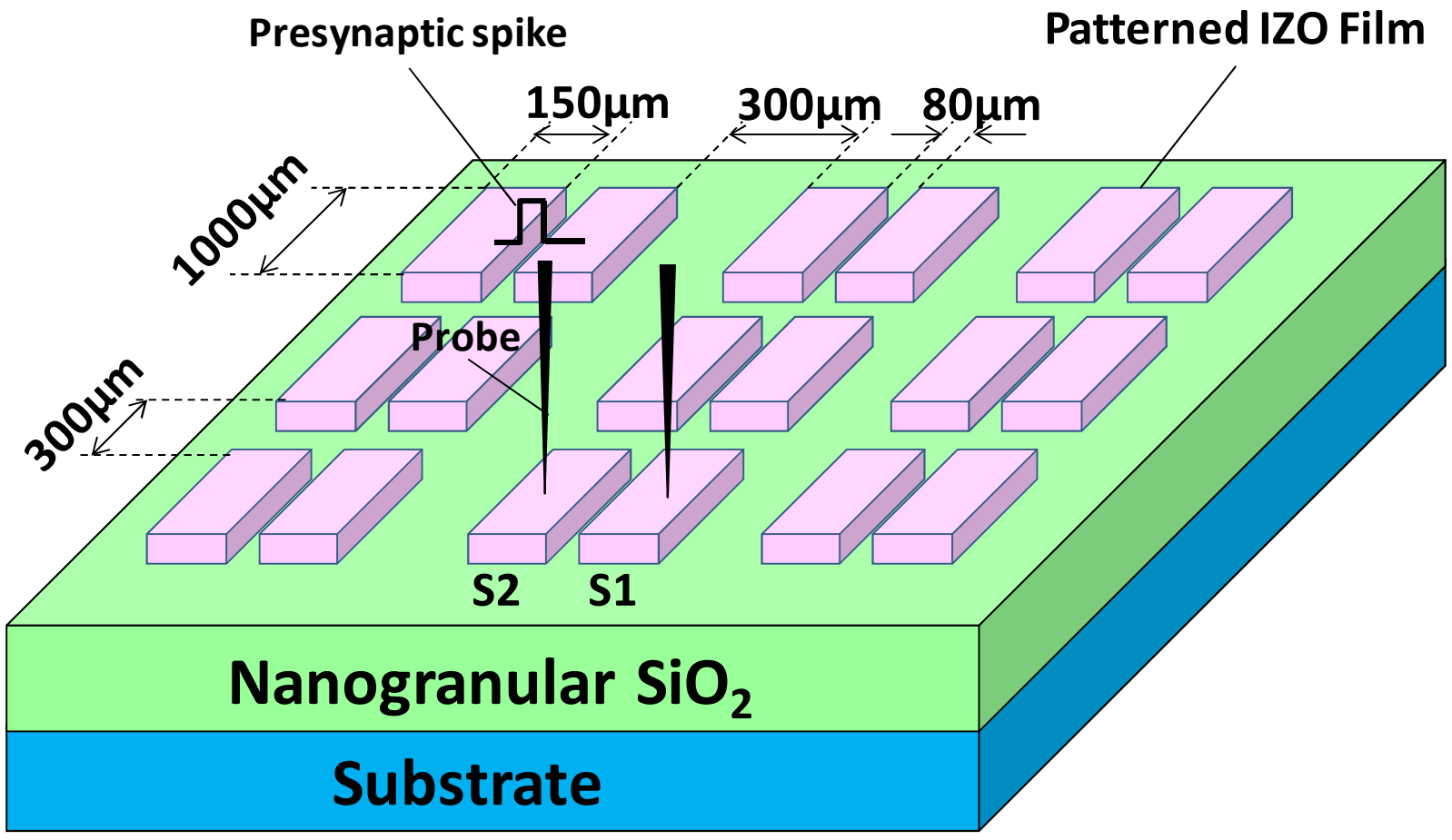

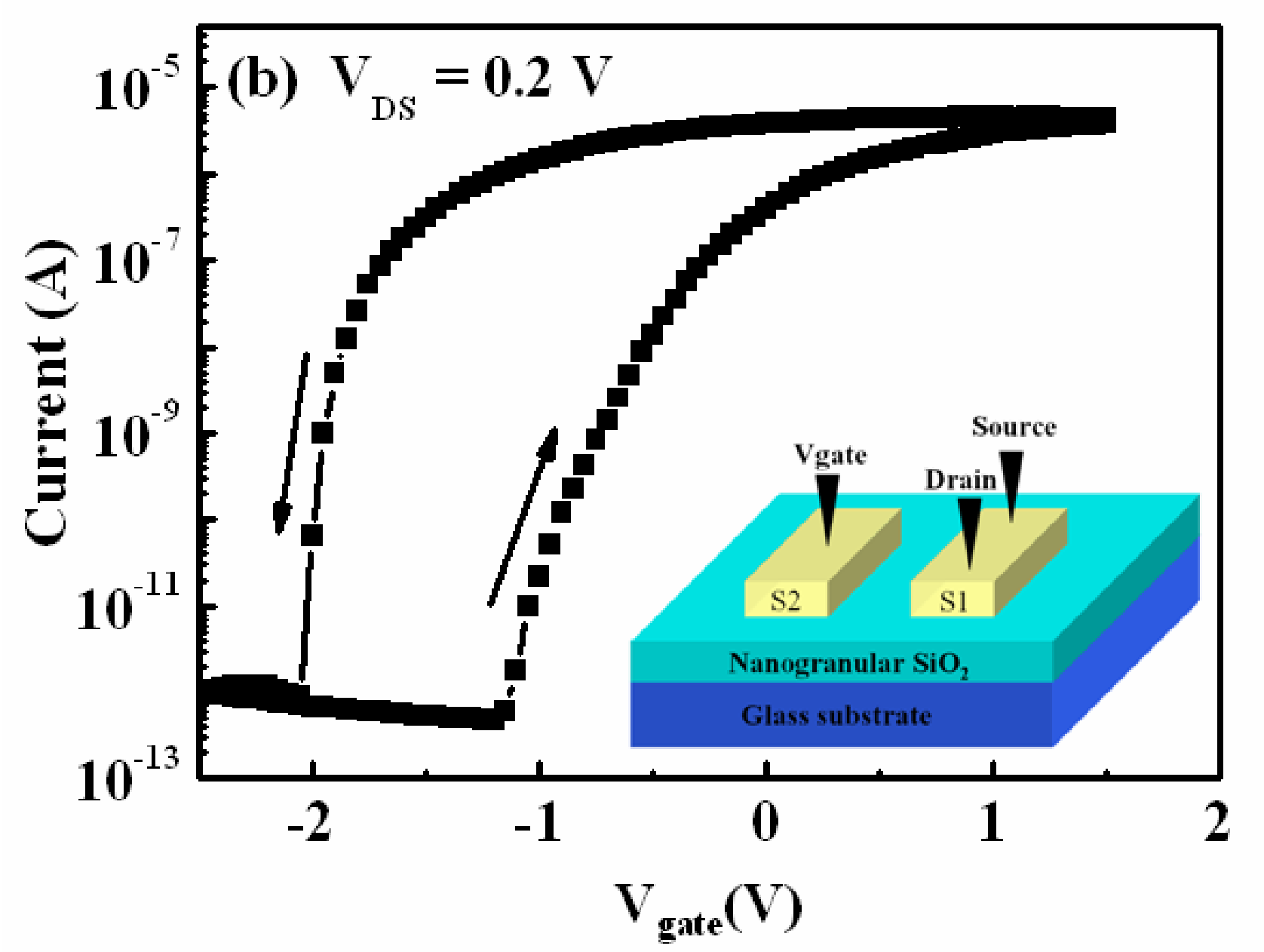

Figure 2. (a) Schematic diagram of the synaptic arrays on nanogranular $SiO_2$ proton conductor film on glass substrates. Each pair of patterned indium-zinc-oxide (IZO) films is a synaptic unit, which can be used as a synapse. (b) Later field-effect modulation of patterned IZO film on nanogranular P-doped $SiO_2$ proton conductor film, Inset is the Schematic diagram for electrical measurement.

Figure 2 (a) shows the schematic diagram of our proposed oxide-based artificial synaptic arrays on nanogranular P-doped $SiO_2$ proton conductor films. P-doped Nanogranular $SiO_2$ films were deposited on glass and Si (100) substrates by PECVD method and 60-nm-thick indium-zinc-oxide (IZO) film arrays were deposited by RF magnetron sputtering and patterned by a nickel shadow mask. Detailed device fabrication process can be found in Experimental Methods. Figure 2 (b) shows the later field-effect modulation of patterned IZO film (S1) by voltage applied on $S_2$, and inset is the schematic diagram for electrical curves measurement. The resistivity of the patterned IZO film can be tuned with about seven magnitudes. A large hysteresis with an anticlockwise window of ~1.0 V was observed in the current-voltage curve due to the high-density mobile proton in the nanogranular P-doped $SiO_2$ proton conductor film. The proton density is estimated to be $6.3\times10^{12}$ /cm$^2$ by $N = \Delta V_{th} C_i / e$, where $\Delta V_{th}$ is the hysteresis window and $C_i$ =1.0 μF/cm$^2$ is the capacitance per unit.

In neurosciences, a potential spike signal in presynaptic neuron can trigger an ion excitatory postsynaptic current (EPSC) or inhibitory postsynaptic current (IPSC) in a postsynaptic neuron through a synapse. [23] This enables the postsynaptic neuron to collectively process the EPSC or IPSC through ~$10^4$ synapses to spatial and temporal correlated functions. The neurotransmission arise from a large number of mechanisms such as synaptic plasticity.[17] Synaptic plasticity is the ability of the connection, or synapse between two neurons to change in strength in response to either use or disuse of transmission over synaptic pathways.[18] Plasticity can be categorized as short-term plasticity (STP), lasting a few seconds or less, or long-term plasticity (LTP), which lasts from minutes to hours[19-21]. Based on the plastic connection of synapses, human memory is created by the dynamic change of neural network. And it is believed that the short term memory (STM) can transform to long term memory (LTM) through a process of rehearsal. [22] Based on these biological theories, each individual 'synapse' of the synaptic arrays was plastic and proved by our experiments. The process of rehearsal induced memory retention change can also mimicked by the individual synapse.

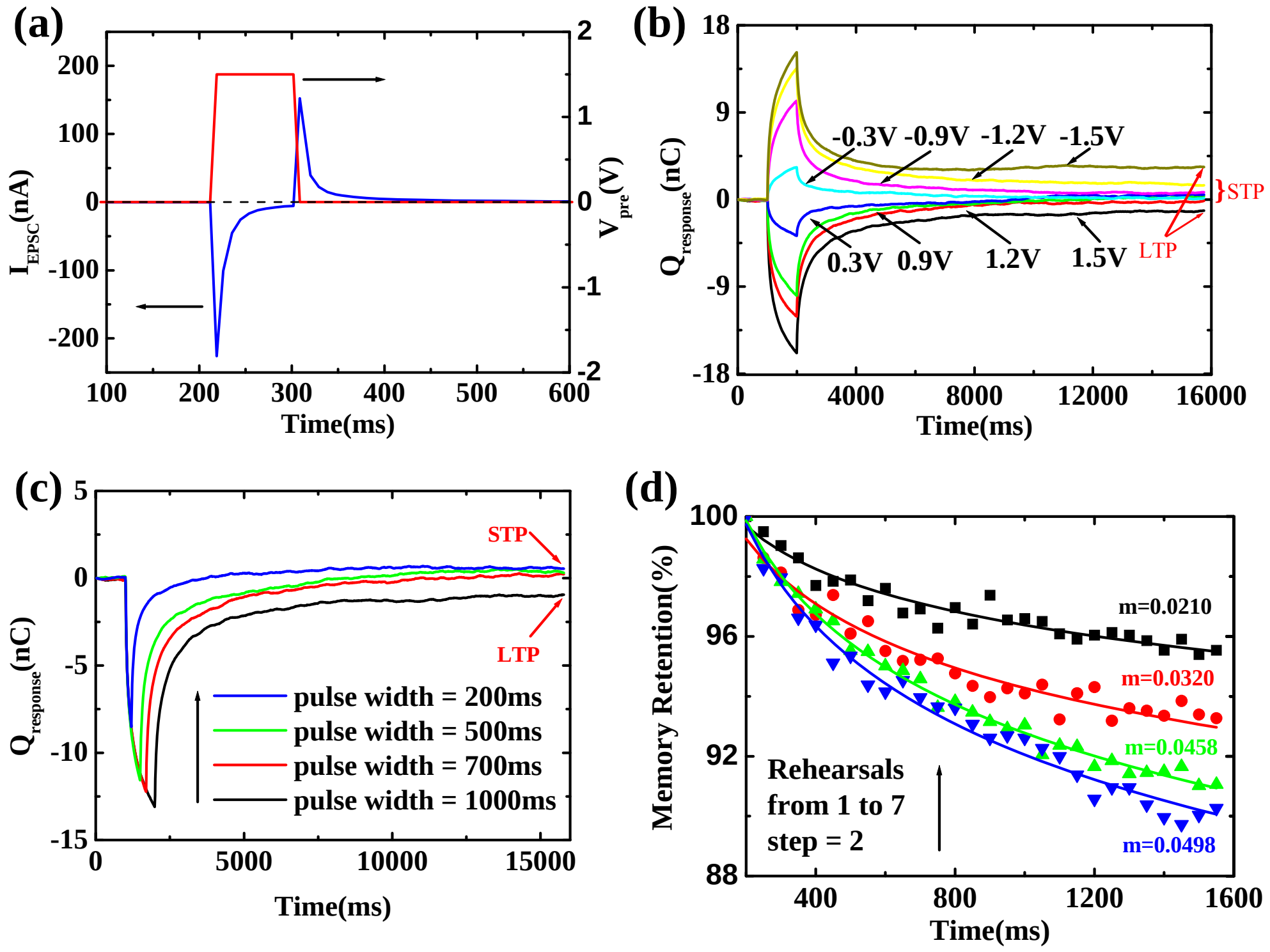

Figure 3. Mimicking of synaptic plasticity. (a) The postsynaptic current. Presynaptic spike (1.5V, 100 ms) was applied on $S_2$, and postsynaptic current was measured on $S_1$. The red line is the spike and blue line is the postsynaptic current. (b) Presynaptic spike amplitude dependent charge retention properties on S1. (c) Presynaptic spike width dependent charge retention properties on $S_1$. (d) STM to LTM transitions by rehearsal. With rehearsals from 1 to 7 and step of 2, the decay rate was decreased. The retention change can be fitted by a power function, $y=b\times t^{-m}$, used to analyze psychological behavior (ref 3). The value of m corresponded each rehearsal is 0.050, 0.046, 0.032 and 0.021, respectively.

---

In order to illustrate the plasticity of individual ‘synapse’, a pair of patterned IZO films ($S_1$ and $S_2$ in Fig 2 (a)) is regarded as a synapse. When a presynaptic spike (1.5V, 100 ms) is applied on $S_2$, excitatory postsynaptic current (EPSC) can be spontaneously measured on $S_1$. Figure 3 (a) shows the time dependent transient current curve measured on $S_1$. The transient current is time dependent and gradually reduces to zero within ~250 ms. It is because when a positive pulse is applied on $S_2$, protons in $SiO_2$ nanogranular film will migrate and accumulate at the interface of $S_1$. After the spike, some protons will diffuse back due to the concentration gradient. The migration of proton would induce a transient current on S1, which result in the EPSC. The negative spikes would have similar effect. Our results are in good agreement with the biological processes of EPSC. [23-25]

In order to investigate the influence of pulse amplitude and pulse width to the synaptic plasticity, we apply a series of pulses with different amplitudes or different widths as the presynaptic spikes on $S_2$ and measure the postsynaptic current on $S_1$. The changes of charge are calculated by the time integral of the transient current. As shown in Fig 3 (b), when the pulse amplitude is less than 1.5 V, charges will decay to zero within a few seconds. When the pulse amplitude $\geqq$1.5V, charges keep a relative high level (>1nC) even after ten seconds. Figure 3 (c) shows the charge retention on S1 generated by 1.5 V pulses with different widths. When the pulse width is less than 1000 ms, charge decay to zero within a few seconds. When the pulse width$\geqq$1000 ms, the charge keep high than 1.0 nC after ten seconds. These results indicate that mimicking of STP/LTP can be achieved through pulse amplitude or width modulation. So the 'synapse' is plastic, and its weight is a function of input pulse amplitude and width.

In order to investigate the influence of pulse number or the rehearsal time to the synaptic plasticity, different numbers (from 1 to 7) of pulses (1.5 V, 200 ms) are applied on S2. 200 ms after the stimulations, the decay of conductance of S1 is measured by applying pulse list, as shown in Fig 3 (d). The measured amplitude of pulse list is set to be 0.01 V in order to minimize the influence to proton migration. The decay curves can be fitted by a power function (m) generally used in psychology, where the ratio of memory retention increases with the number of rehearsals. [8] In our experiment, the decay rate or the value of *m* decreases from 0.05 to 0.02 when the number of pulse increases from 1 to 7. Therefore, if the pulse number defined as the repetition time and the conductance change of S1 defined as the memory retention. With more experiments, the device can also mimic the spike timing dependent plasticity (STDP), an essential learning/memory laws of synaptic function, [23] by measure the conductance change of the patterned IZO film (see support information 1).

If connected by hard wires, n(n-1)/2 wires are needed to build the interconnection of artificial neurons networks, so it would be too enormous and complex to accomplish mimicking of the $10^{15}$ connections as in our brain. It is interesting to find that our synaptic arrays are able to mimic some transmission behaviors of neural networks without any intentional hard-wired connection due to the electric-field induced lateral proton modulation. Spikes with same amplitude of 1.5 V and same width of 200 ms are applied on presynaptic patterned IZO films with different distance to S1, and the EPSC is measured on $S_1$. The changes of absolute value of maximum EPSC ($I_{max}$) with distance is shown in Fig. 4 (b). The $I_{max}$ can be fitted by a function of distance between two patterned IZO films (d), that's:

$$I_{max} = \kappa \bullet d^{\,n}.$$

κ is a constant at a certain amplitude of the pulse, and n is a factor reflected the attenuated characteristic of the power function. The parameters of $\kappa=1.682 \cdot 10^{-8}$ and n=-0.3485 are fitted from the experiment data (blue line in Fig 4 (b)). The factor n with small absolute value indicates the characteristic of weakly correlation with distance.

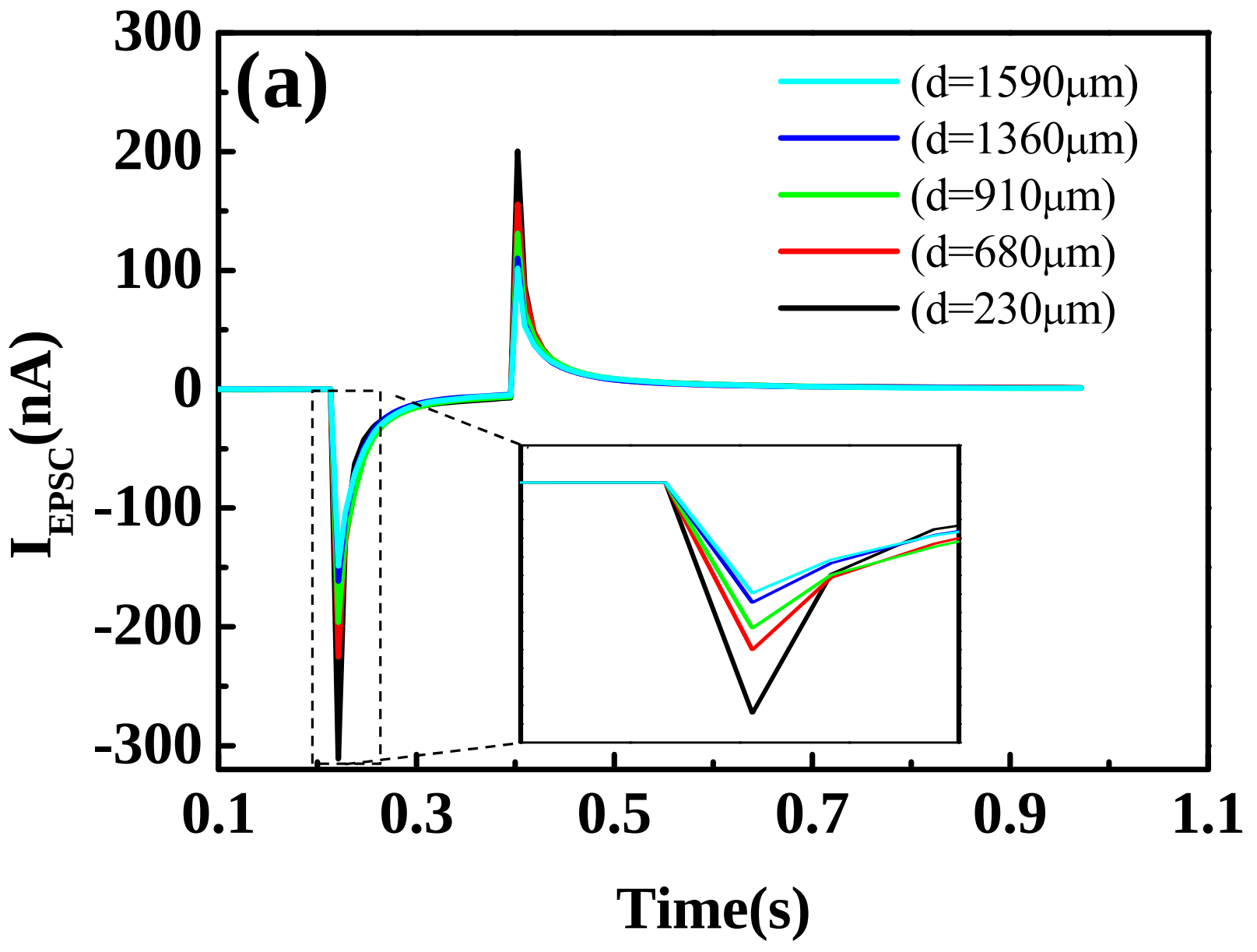

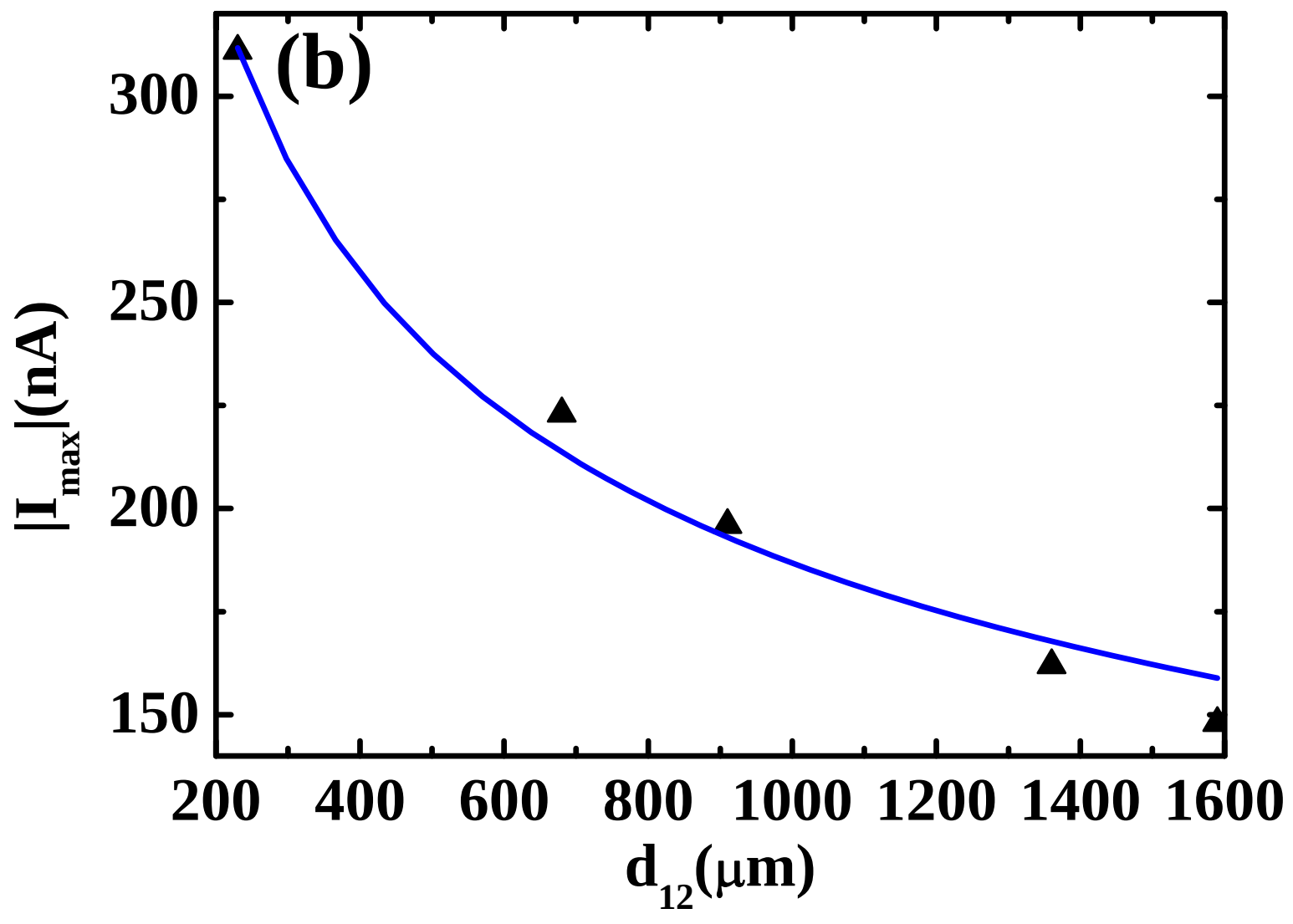

Figure 4. Weak correlation of EPSC with distance of the artificial synaptic arrays. (a) The postsynaptic current generated by presynaptic spikes (1.5 V, 200 ms) with different distances. Inset shows a magnified view of the maximum EPSC. (b) distance-dependent maximum EPSC of the artificial synaptic arrays. The triangle dots were the experiment data and it can be fitted by a power function (the blue solid line).

---

The electric-double-layer effect of the nanogranular P-doped $SiO_2$ proton conductor film also plays a crucial role in mimicking neural network. Figure 4 (a) shows the distance dependent EPSC current of the artificial synaptic arrays. After stimulation, all EPSC currents reduce to zero within almost same time (~200 ms), which is comparable with those observed in biological synapses. [23-25] Long time constants are due to the high specific capacitance of EDL. Therefore, the synaptic arrays can potentially be fabricated for an interconnected neuromorphic system on large scale with more flexible and feasible than hard-wires.

One essential function of neural network is to integrate large numbers of inputs that arrive on a certain dendritic trees and react properly, as shown in Fig 5 (a). [26] In other words, the summations determine whether or not an action potential will be triggered by the summation of postsynaptic potentials. Spatial summation as one form of summation was derived from Katz's finding in 1951, which indicated that action potential generation can be triggered by the summation of individual units (figure 5 (a)). [27] To demonstrate the networked behaviors of the array, the device was used to mimic spatial summation of neurons. As shown in Fig. 5 (b), we use four patterned IZO films, one ($S_1$) for dendrite (or cells body) and three ($S_2$, $S_2$, $S_4$) for separate axon terminal. The presynaptic spikes were applied on $S_2$, $S_3$ and $S_4$, and the EPSC was measured on $S_1$. As shown in Fig. 5 (c), when only a 0.5 V pulse is applied on one "axon terminal" ($S_2$), a max negative current of 49.5 nA (green line) is measured on $S_1$. When three 0.5V pulses are applied on three "axon terminal", a max negative current of 158 nA (green line) can be measured on $S_1$. When a 1.5 V pulse is applied on single "axon terminal" ($S_2$), a max negative current of 165 nA (red line) is measured on $S_1$. On the contrary, as shown in Fig. 5 (d), when only applying 0.75V pulse on $S_2$, a max positive current of 77.5 nA (red line) was measure on $S_1$. A very low max negative current of 18.6 nA is measured on $S_1$when $S_2$=1.5 V and $S_3$=$S_4$=-0.75 V. The above results indicate the pulsed signals of three axon terminals can be summed, corresponding to summation behaviors in vivo.

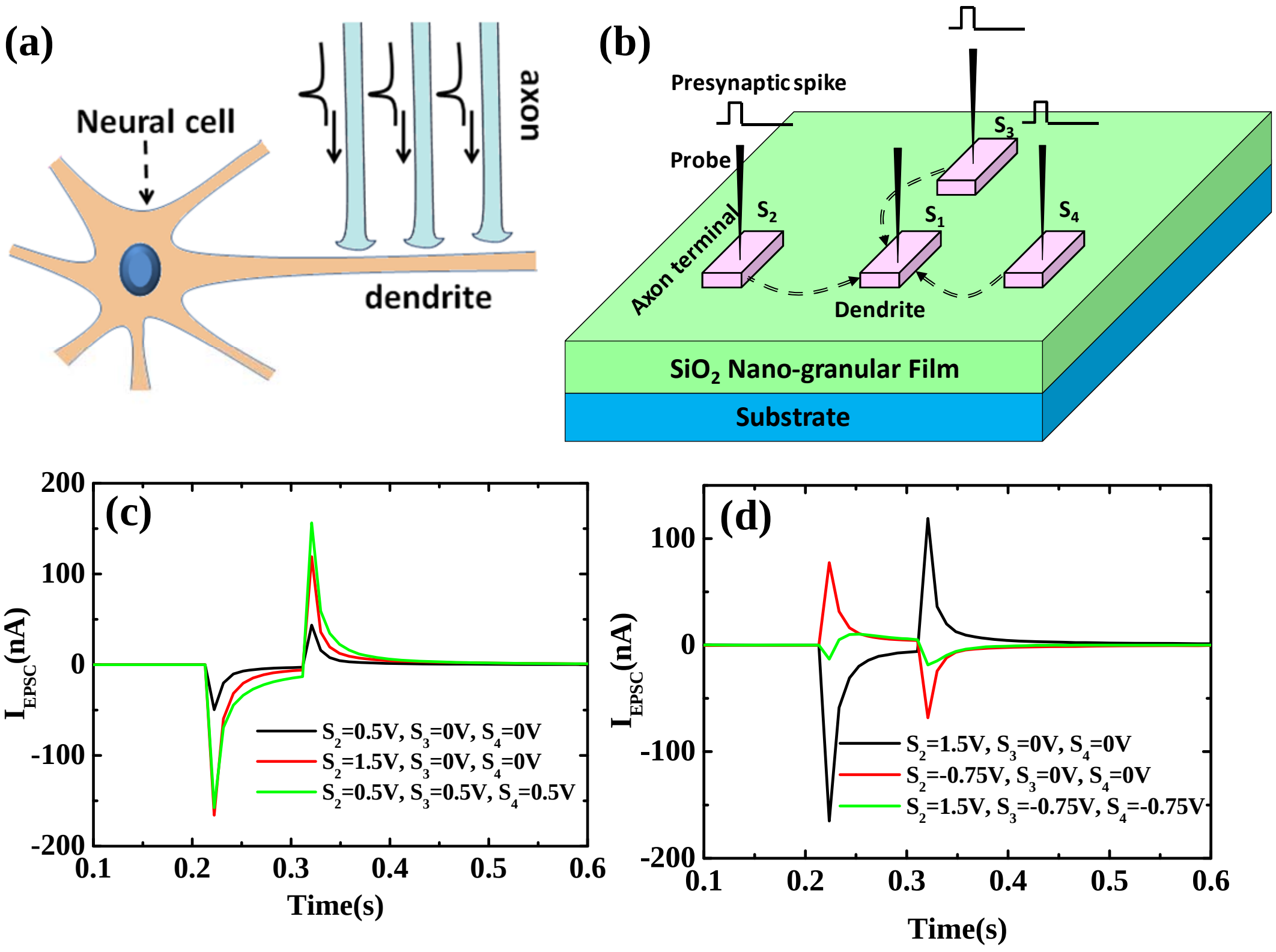

Figure 5.Mimicking of the spatial summation behaviors of the artificial synaptic arrays. (a) Schematic diagram of spatial summation behavior in vivo. (b) Schematic diagram of mimicking spatial summation behavior in artificial synaptic arrays. Three patterned IZO films ($S_2$, $S_3$, $S_4$) are used as axon terminal and one ($S_1$) as dendrite. The presynaptic spikes were applied on the three 'axon terminal', and the summation was measured on the 'dendrite'. Distance from the $S_2$, $S_3$ and $S_4$ to $S_1$ are 680 μm, 1320 μm and 680 μm, respectively. (c) and (d) Spatial summation functions of the artificial synaptic arrays with different amplitudes of presynaptic spike applied on $S_2$, $S_3$ and $S_4$.

---

B.W. Mel gives a simple formula to describe summation process: [28]

$$y = \sum_{j=1}^{m \ branches} g\left(\sum_{i=1}^{k \ synapses} \omega_{ij}\chi_{ij}\right)$$

$\omega_{ij}$ is the weight of axon terminal i from branch j, and $\chi_{ij}$ is the amplitude of spike, y is neuron's overall output value and g is typically modeled as a sigmoid or power function. Combining with our impedance model, the parameter d can be axon terminal's weight correlative to distance, and the κ can represent the amplitude of spike. So mimicking m axon terminals' summation on neuron i of our arrays can be described by a mathematical formula (in order to simplify, neuron i was assumed with one branch or dendrite):

$$y_i = \sum_{j=1}^{m\ synapses} \kappa_{ij} d_{ij}^{\ n}$$

Output $y_i$ is represented by transient current of the dendrite patterned IZO film. The formula is matched with the structure of power function and can be applied to calculate the impact of summation behavior mentioned above (like the slightly distinction mentioned in figure 4c). Because the transient current can be summed, the array can also be used to achieve some logical function if the location of the patterned IZO film was proper designed.

In conclusion, oxide-based artificial synaptic arrays on nanogranular P-doped $SiO_2$-based proton conductor films were fabricated at room temperature. Some plasticity behaviors of a synapse, like STP, LTP and STDP were mimicked on an individual synaptic device. Without any intentional hard-wire connection, such synaptic arrays were intercoupled due to the electric-field induced proton modulation. The natural interconnection was weakly correlative with distance, and paralleled summation was also experimental demonstrated. These results indicated that without complex hard-wired connections, artificial synaptic arrays on inorganic nanogranular proton conductor film may provide a new approach for building future brain-like computational systems.

## Experimental Methods

All fabrication processes of oxide-based synaptic arrays were performed at room temperature. First, 680-nm-thick P-doped nanogranular $SiO_2$-based proton conductor films were deposited on glass substrate by plasma-enhanced chemical vapor deposition (PECVD) using $SiH_4$ (95%)/$PH_3$ (5%) mixed gases and $O_2$ as reactive gases at room temperature. The flow rates of $SiH_4$/$PH_3$ mixed gases and $O_2$ were 10 sccm and 60 sccm, respectively. The deposition pressure, work power, and deposition time were 30 Pa, 100 W, and 45 min, respectively. Then, 60-nm indium-zinc-oxide (IZO) films were deposited on P-doped $SiO_2$-based proton conductor by magnetron sputtering and patterned with a nickel shadow mask. The IZO film deposition was performed using an IZO target (90 wt% $In_2O_3$ and 10 wt% ZnO) with a power of 100 W, a working pressure of 0.5 Pa in Ar ambient with trace of oxygen. The plane size of the patterned ITO films is 150 μm ×1000 μm.

For cross-section scanning electron microscopy (SEM) characterization, $SiO_2$ nanogranular film was deposited on Si (100) substrate. For transmission electron microscopy (TEM) characterization, very thin $SiO_2$ nanogranular film was deposited on TEM Cu grid. The proton conductivity of the $SiO_2$ nanogranular film was determined from Cole-Cole plots measured by Solartron 1260 impedance analyzer at a 50% relative humidity (RH). The Cole-Cole plot consisted of a single semicircle and the proton conductivity was obtained from the intersecting point of the semicircle with the real axis. The proton conductivity, σ, was calculated from R and pellet dimension, i.e., $\sigma=L/(R-R_0)A$, [29] where L, A and $R_0$ are thickness of the film, electrode surface area and the rig short circuit resistance, respectively. The proton conductivity of the nanogranular $SiO_2$ film is calculated to be $3.12\times10^{-4}$ S/cm ($R_0$=30 Ohm). All the electrical measurements of synaptic devices were done by using Keithley 4200 semiconductor parameter analyzer at 50% RH.

Support information 1

STDP

As shown in Fig. s1, pre- and postsynaptic spike pairs (1.5V, 100 ms) are applied to S2 and S1, respectively. Time interval (△t) between pre and postsynaptic from -70 ms to 70 ms were set and it's comparable to biological experiment. [1] Fig. s2 shows one period of the spike pair. The width of stimulation period is 500 ms, and at each period's initiate, there is 100 ms's pause. The spike pair was applied on S2 for 40 repetitions (20 s). Conductance change of the S1 was measured by applying a read pulse with width of 50ms and amplitude of 0.01 V on S1, before and after the 40 repetitions. Fig. s3 shows the STDP data points. The conductance change was represented by relative change of current (△$I_{ss}/I_{s0}$), which shows a function of △t. These results are similar to what has been observed in biological synapses. [1]

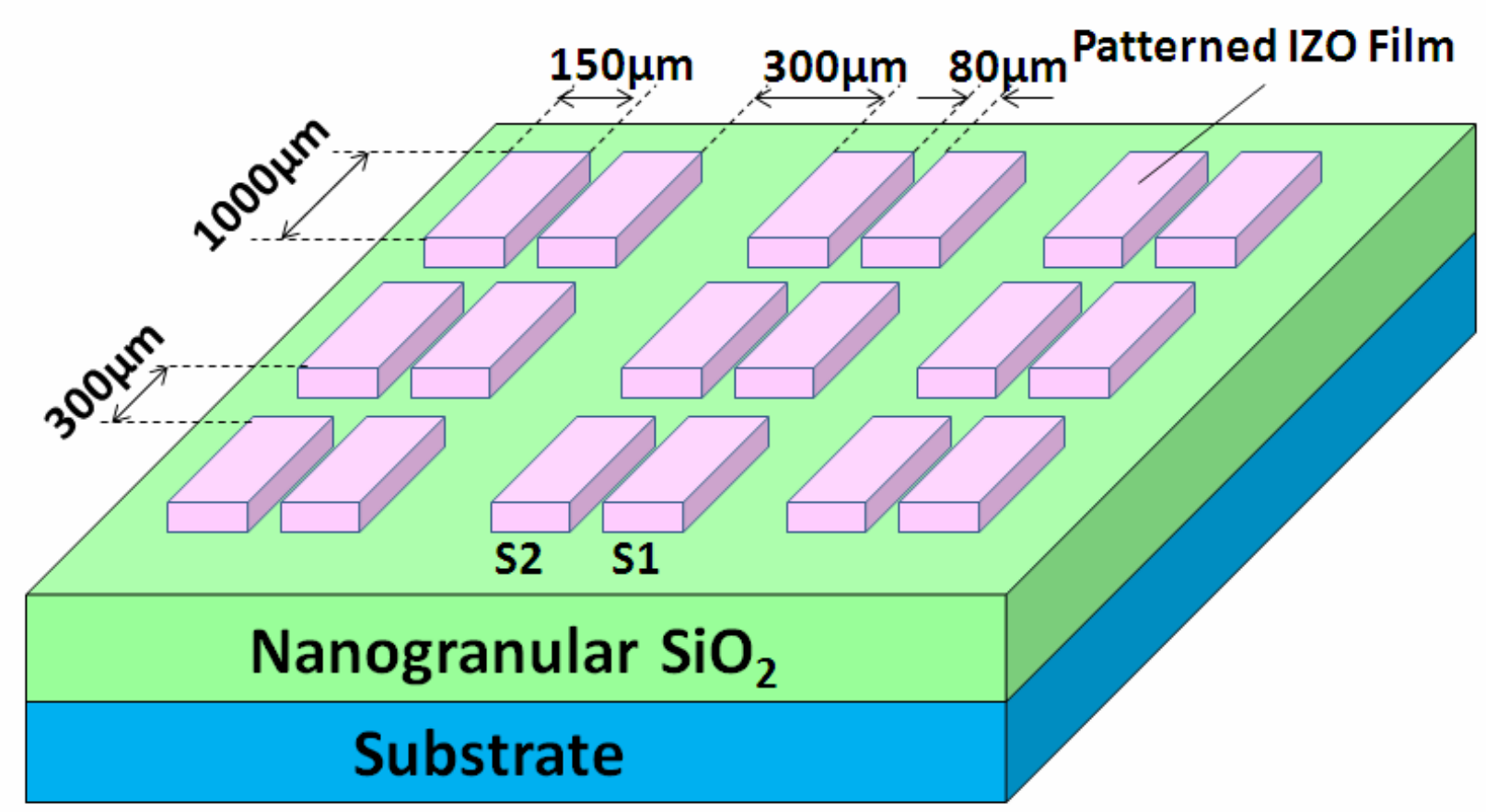

Figure s1. The schematic diagram for mimicking spike-timing-dependent-plasticity.

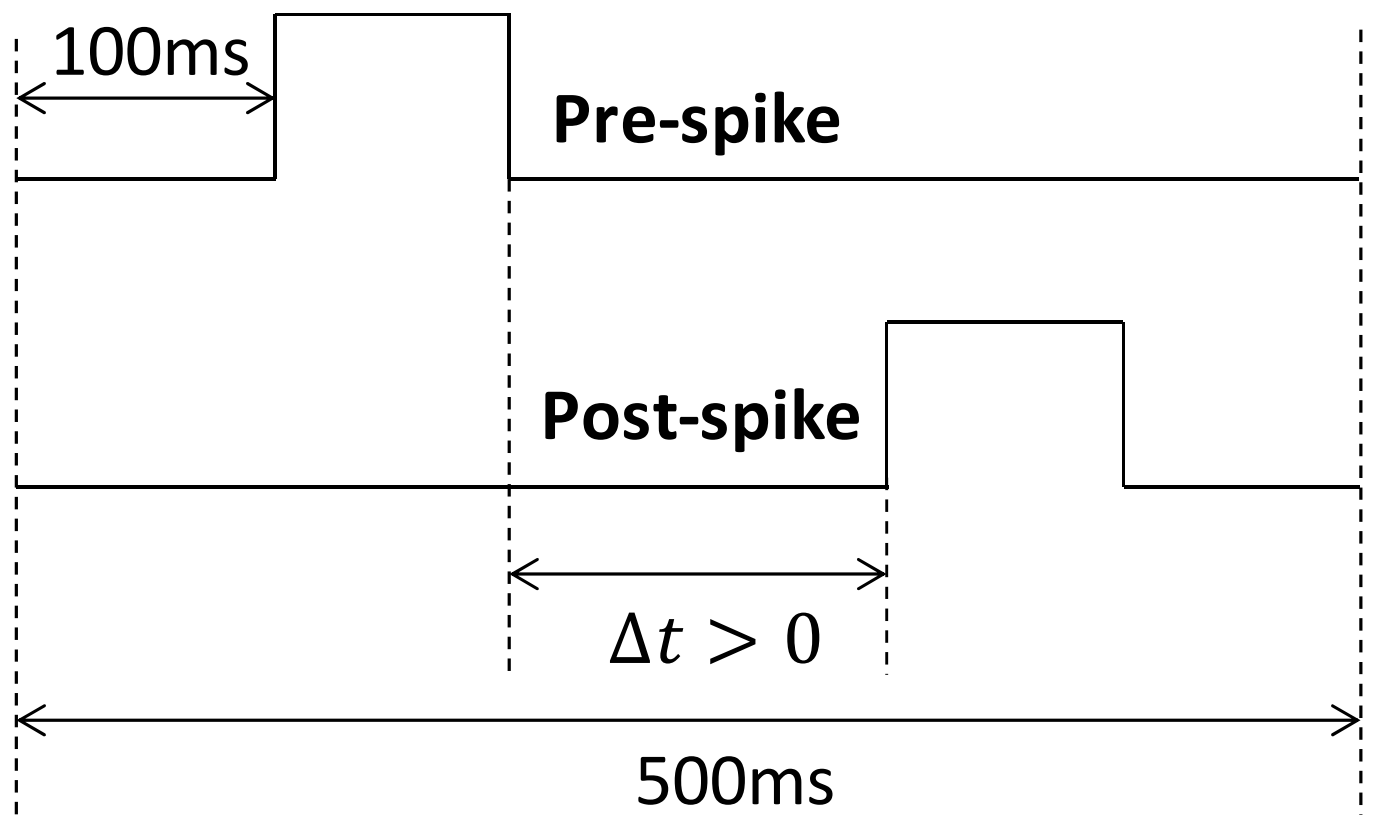

Figure s2 Stimulation period of the pre- and postsynaptic spikes. At each period's initiate there is 100ms's pause. The spikes with amplitude of 1.5V and width of 100ms were applied on $S_2$ and $S_1$.

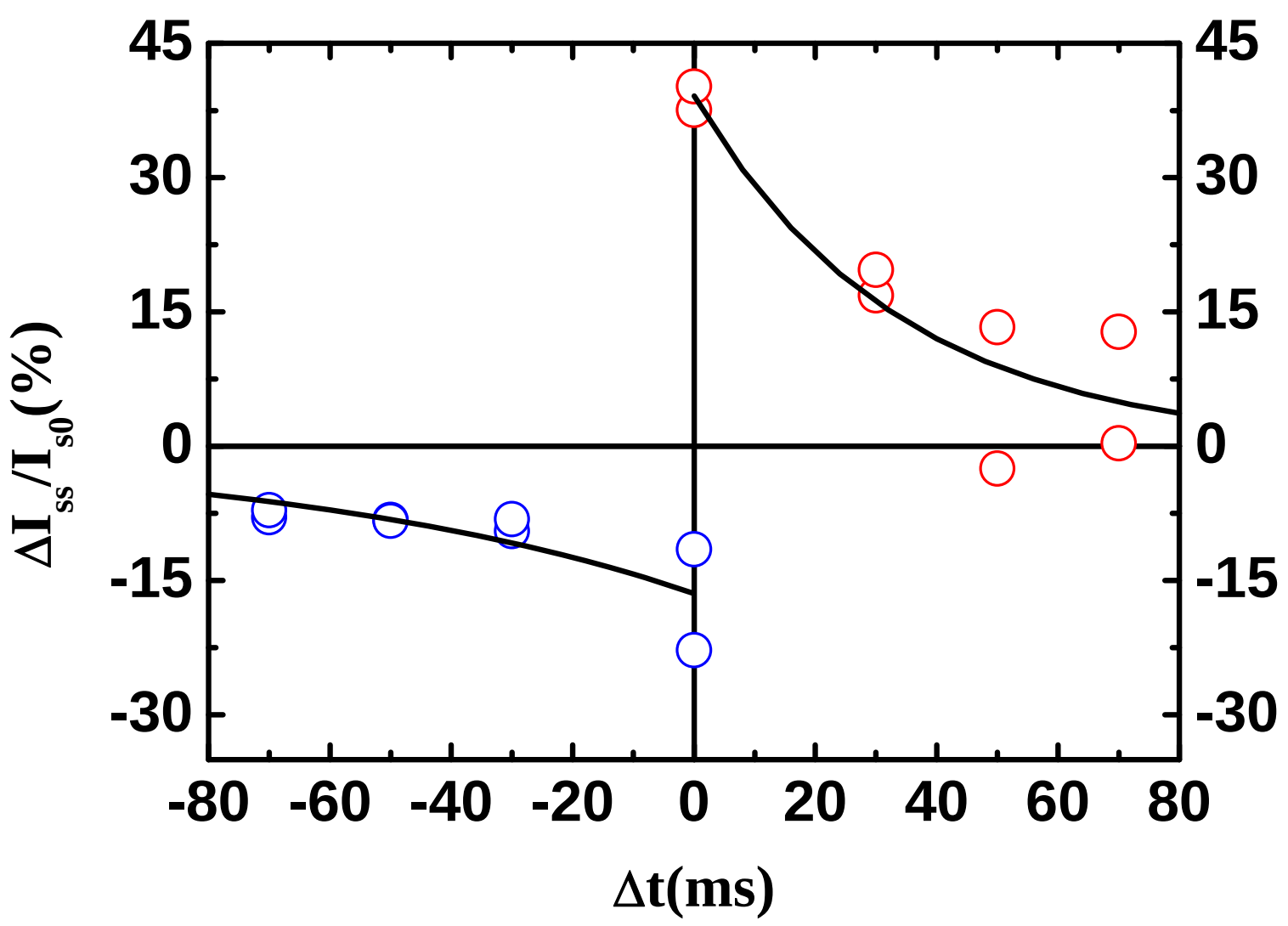

Figure s3 The synaptic weight ($\Delta I_{ss} \ / \ I_{s0}$) versus the spiking time interval $\Delta t$

Support information 2

Derivation of EPSC's peak value

We use the ideal circuit model to derivation the peak value of EPSC. The spike pulse we apply is square wave, which can be described by sum of step function ($u(t)$), that's $U(t) = A \cdot (u(t) - u(t - t_1))$, $A$ is the amplitude of the pulse, we assume the pulse is begun at $t = 0$ and ended at $t = t_1$. All the elements are used Laplacian model (figure s4), and we can obtain two equations:

$$\begin{cases} U(s) = U_R(s) + U_C(s) \\ I(s) = I_R(s) = I_C(s) \end{cases} \tag{1}$$

Then, the $I(s)$ can be rewritten by function of $U(s)$, that's:

$$I(s) = \frac{\frac{1}{R}s}{\frac{1}{RC} + s} U(s) \tag{2}$$

As $U(s) = A \cdot (\frac{1}{s} - \frac{1}{s} e^{-t_1 s})$, we can get the transient current function by inverse Laplace transform:

$$I(t) = \frac{A}{R}(u(t)e^{-\frac{t}{RC}} - u(t - t_1)e^{-\frac{t-t_1}{RC}}) \tag{3}$$

So $I_{max} = \frac{A}{R}$, and the $R \propto d^n$ (if the resistor is linear $n = 1$), then the $I_{max}$ can be described by the equation:

$$I_{max} = \kappa \cdot d^n \tag{4}$$

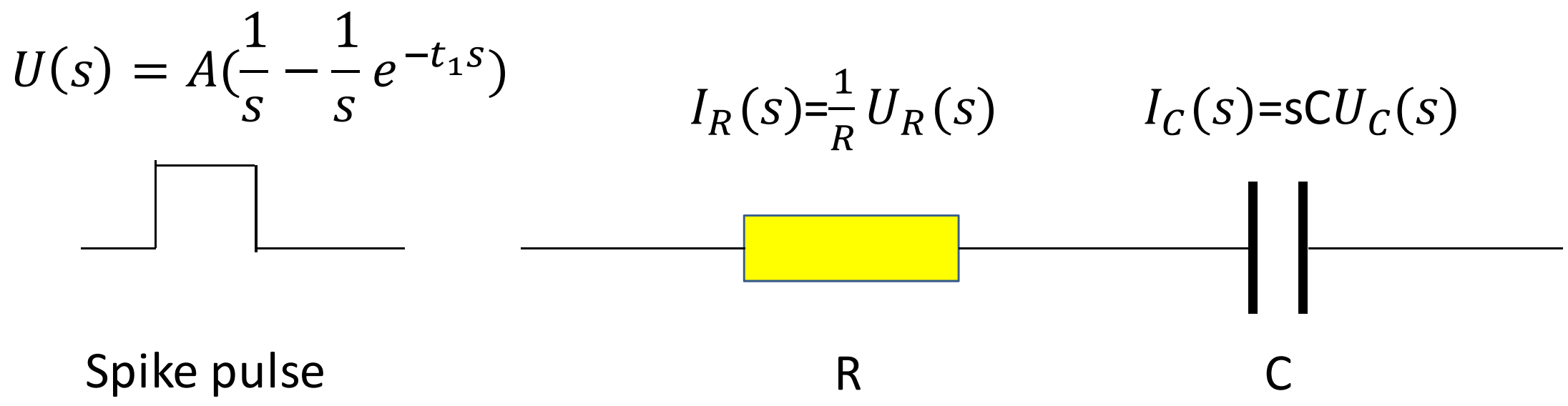

Figure s4 Laplacian model of each element